\documentclass[prd,11pt,aps,superscriptaddress,floatfix,showkeys,showpacs]{revtex4}
\usepackage{graphicx,amsmath,bm,array,subfigure}
\usepackage[linktocpage=true,colorlinks=true,linkcolor=blue,citecolor=blue]{hyperref}

\usepackage{slashed}
\usepackage{lmodern}
\usepackage {amssymb}

\begin{document}
\title {Heavy quark complex potential in a strongly magnetized hot QGP medium}
\author{Balbeer Singh }
\email{balbeer@prl.res.in}
\affiliation{Theory Division, Physical Research Laboratory,
	Navrangpura, Ahmedabad 380 009, India}
\affiliation{Indian Institute of Technology Gandhinagar, Gandhinagar 382355, Gujarat, India}
\author{Lata Thakur}
\email{latathakur@prl.res.in}
\affiliation{Theory Division, Physical Research Laboratory,
	Navrangpura, Ahmedabad 380 009, India}
\author{Hiranmaya Mishra }
\email{hm@prl.res.in}
\affiliation{Theory Division, Physical Research Laboratory,
	Navrangpura, Ahmedabad 380 009, India}
\begin{abstract}
We study the effect of a strong constant magnetic field, generated in relativistic heavy ion collisions, on
the heavy quark complex potential. We work in the strong magnetic field limit with the lowest Landau level
approximation. We find that the screening of the real part of the potential increases with the increase in the
magnetic field. Therefore, we expect less binding of the $ Q\bar{Q} $ pair in the presence of a strong magnetic field.
The imaginary part of the potential increases in magnitude with the increase in magnetic field, leading to an
increase of the width of the quarkonium state with the magnetic field. All of these effects result in the early
dissociation of $ Q\bar{Q} $ states in a magnetized hot quark-gluon plasma medium.
\end{abstract}
\pacs{11.10.Wx,14.40.Pq,12.38.Bx,12.38.Mh}
\keywords{Heavy quarkonium, Heavy quark-antiquark potential, Dielectric permittivity, 
	Decay width, Debye mass, String tension.}

\maketitle 
\section{Introduction}
\noindent
Relativistic heavy-ion collisions (RHIC) at Brookhevan and LHC at CERN provide a lot of information regarding 
the deconfined state of matter known as quark-gluon plasma (QGP). In recent studies ~\cite{Kharzeev:2007jp} it has been realized that noncentral collisions also produce a  strong magnetic field in the
direction perpendicular to the reaction plane which has a number of interesting phenomenological consequences~\cite{Tuchin:2013ie, Kharzeev:2007jp, Tuchin:2010vs,Tuchin:2010gx,Mohapatra:2011ku}.
The estimated strength of the magnetic field is 
$B =m_{\pi}^2=10^{18}$ Gauss  at the RHIC energies and 
$B =15m_{\pi}^2=1.5 \times10^{19}$ Gauss at the LHC energies~\cite{Kharzeev:2007jp,Skokov:2009qp},
where $ m_{\pi} $ is the pion mass. Possibly, such a huge magnetic field might have existed in the early Universe as an origin of the present large scale
cosmic magnetic field~\cite{Vachaspati:1991nm,Grasso:2000wj}.

The medium in the presence of the magnetic field requires the modification of the existing theoretical tools that can be used  to investigate the various properties of QGP. 
How the magnetic field modifies the properties of strongly interacting matter has been the subject of
various theoretical efforts~\cite{Kharzeev:2012ph,Miransky:2015ava}.
Various studies, based on
direct numerical investigations using lattice QCD~\cite{DElia:2010abb,DElia:2011koc,Bali:2012zg} %[13-17]Bali:2011qj
and effective theoretical investigations using different methods
including, e.g., perturbative QCD studies, model
studies, and anti-de Sitter/conformal field theory correspondence studies~\cite{Alexandre:2000jc,Agasian:2008tb,Fraga:2008qn,Mizher:2010zb,Gatto:2010qs,Gatto:2010pt,Andersen:2011ip,Erdmenger:2011bw,Gorbar:2011jd,Skokov:2011ib,Fraga:2012fs,Fraga:2012ev,Fraga:2012rr,Andersen:2012dz,Preis:2012fh,Ferrari:2012yw,Fayazbakhsh:2012vr,Fukushima:2012xw,dePaoli:2012cz,Kojo:2012js,Kojo:2013uua} 
have predicted many
interesting phenomena affecting the properties of strongly
interacting matter in the presence of strong magnetic
backgrounds.
It is still not certain though to what extent such
phenomena will be detectable in heavy ion experiments.
In this aspect, effects regarding the physics of heavy quark bound states
($ Q\bar{Q} $) are of particular interest, since they are more
sensitive to the conditions taking place in the early stages.
They have large masses and are not affected by the thermal medium.
Thus, the heavy quarkonium states
are considered as one of the powerful probes to study
the deconfining properties of the strongly interacting medium.
 
Matsui and Satz~\cite{Matsui:1986dk} predicted
a suppression of the $ J/\psi $, being caused by the shortening of
the screening length for color interactions in the QGP.

Initially, it was thought that the large magnetic field produced in such collisions decays very fast~\cite{McLerran:2013hla}. %$ t_Q\simeq 0.2 $ fm
 However later, it has been argued that in a medium with finite electric conductivity can sustain strong B-field even in the QGP phase as discussed in Refs.~\cite{Tuchin:2010gx,Das:2017qfi}. %[PRC 83, 017901 (2011),PRC 96, 034902 (2017)]. 
The reason being that the magnetic field does not decay very rapidly due to the induced currents in the medium and satisfies a diffusion equation. Indeed, it is argued that ~\cite{Tuchin:2010gx,Das:2017qfi} magnetic field decreases fast initially, but at later times, matter effects become more important slowing down the decrease of the magnetic field significantly. This effectively leads to the external magnetic field to be a slowly varying function of time during the entire QGP lifetime.

Simultaneously heavy quark and antiquarks pairs also develop into a 
physical resonances over a formation time $ t_{form} \sim 1/E_{bind}$,
e.g.,
the $ c\bar{c} $ pairs form
resonances at $ t_{c\bar{c}} \sim 0.3 $~fm. 
 
Therefore, it is reasonable to assume that the $ Q\bar{Q} $ pair is strongly influenced by the magnetic field. 
As regards to the effects more directly related to
color interactions, various studies have considered the
possible influence of an external magnetic field on the
static quark-antiquark potential~\cite{Bonati:2015dka,Bonati:2016kxj} and the screening masses~\cite{Bonati:2017uvz}. So far, different studies
have been carried out on the effects of magnetic field for static
properties of quarkonia~\cite{Alford:2013jva,Marasinghe:2011bt,Cho:2014exa,Guo:2015nsa,Bonati:2015dka,Rougemont:2014efa,Dudal:2014jfa,Sadofyev:2015hxa} and of open heavy flavors
~\cite{Machado:2013rta,Machado:2013yaa,Gubler:2015qok,Fukushima:2015wck,Das:2016cwd}.
The effect of magnetic field on quarkonium production has been discussed in Refs.~\cite{Machado:2013rta,Guo:2015nsa}. Further, the influence of strong magnetic field on the evolution of $ J/\psi $  and the magnetic conversion of $ \eta_c $ into $ J/\psi $ has been discussed in Refs.~\cite{Marasinghe:2011bt,Yang:2011cz}. %and~\cite{Yang:2011cz} . 
Since masses of heavy quarks ($ m_Q $) are much larger than QCD scale ($ \Lambda_{QCD} $),
the velocity $ v $ of heavy quarks in the bound state is small, and
the binding effects in quarkonia at $ T=0 $ can be
understood in terms of nonrelativistic potential models such as Cornell potential which can be derived directly from QCD using the effective field theories (EFTs)~\cite{Eichten:1979ms,Lucha:1991vn,Brambilla:2004jw}.

In this work we investigate the effects of the magnetic fields on the properties of quarkonium states. Quarkonium states are well understood to be a $ Q\bar{Q} $ pair bound by the Cornell potential which is the sum of the Coulombic potential induced
by a perturbative one-gluon exchange at short range 
and linearly rising potential at large
distances~\cite{Eichten:1974af}.
In recent years, the quarkonium properties have been investigated by correcting both the Coulombic and string terms of the heavy quark potential through the perturbative HTL dielectric permittivity in both the isotropic and anisotropic medium in the static~\cite{Agotiya:2008ie,Thakur:2012eb,Thakur:2013nia,Kakade:2015xua,Agotiya:2016bqr} as well as in moving medium~\cite{Thakur:2016cki}.  
In the present investigation, we focus on the modification of the heavy quark potential in the presence of strong magnetic field.

For the magnetic field considered in the present work
, $ m_Q $ is still the largest scale and the requirement
$m_Q \gg T $ and  $m_Q \gg \sqrt{eB} $ is satisfied, so the heavy quark potential can still be described by a nonrelativistic potential.
 In order to incorporate the magnetic field effect in the heavy quark potential we first calculate the fermionic contribution to
the gluon self-energy in the presence of magnetic field in imaginary time formalism by using the Schwinger propagator~\cite{Schwinger:1951nm} and calculate the Debye screening mass.
Since gluons do not interact with the external magnetic field, the magnetic field dependent contribution arises from the quark loop only. Thus, the total gluon self-energy will be the sum of the gluon (T dependent) and quark (B dependent) loop. As in the deconfined states the gluonic degrees of freedom are larger than the quark degrees of freedom so one must take  both of the effects into account to estimate the Debye mass even in the presence of the magnetic field. 

In the strong
magnetic field approximation only the lowest Landau levels (LLL), which are the energy levels of a moving charged particles, remain active, and the LLL dynamics become important~\cite{Gusynin:1995nb}. Therefore, we have taken only the LLL contribution in our calculation. 
This is a reasonable approximation as quarkonia are produced during the initial stages of collision when the magnetic filed is rather large.
Our aim here is to find a field-theory motivated parametrization of the potential that depends both on the temperature and the magnetic field, which will be
identified through the Debye screening mass, $ m_D $. Further, in the presence of constant magnetic field, the resulting heavy quark potential is expected to be anisotropic. However, as we shall demonstrate later, the anisotropic contribution vanishes in the limit of massless light quarks. In this limit the potential remains still isotropic. Here we follow  Ref.~\cite{Burnier:2015nsa} where the potential is modified by bringing together the generalized Gauss law with the characterization of in-medium effects through the dielectric permittivity.
We demonstrate the effect of a strong homogeneous magnetic field on the Debye screening and Landau damping induced thermal width obtained from the imaginary part of the quark-antiquark potential.   
In the medium, both the perturbative (Coulombic) and nonperturbative (string) part of the heavy quark potential get modified. As the potential is a complex quantity ~\cite{Laine:2006ns,Dumitru:2009fy,Dumitru:2010id,Thakur:2013nia,Thakur:2016cki,Rothkopf:2011db,Margotta:2011ta,Burnier:2009yu} due to which an additional complication will arise. Therefore, a significant description of a relevant physics must illustrate both the effects of Debye screening
in the real part of the potential and  Landau-damping induced thermal width obtained from the imaginary part of the potential.
Recently, the heavy quark potential in a static and strong homogeneous magnetic field has been studied in \cite{Hasan:2017fmf}. However, the authors have studied the effect of magnetic field on the real part of the potential only and they have neglected the gluonic contribution in the Debye mass whereas we have included the gluonic contribution also as discussed above.

This paper is structured as follows. In Sec.~\ref{I} we  first calculate the gluon self-energy using the imaginary time formalism and then calculate the Debye mass by taking the static limit of self-energy. In Sec.~\ref{II} we discuss the heavy quark complex potential in the presence of magnetic field and shows the variation of the
real and imaginary parts of the potential with the various
values of magnetic field at different temperatures. In Sec.~\ref{III} 
we investigate the decay width and discuss its variation with the magnetic field for both bottomonium and charmonium ground states. Finally, we present our
conclusion in Sec.~\ref{IV}. 
\section{Gluon self-energy in LLL in the presence of magnetic field}
\label{I}
In this section we evaluate the one loop gluon self-energy arising from quark loops which gets modified in the presence of
the magnetic field. We shall use it to estimate the Debye  screening mass for the gluons defined through the vanishing four momentum of the polarization tensor. Moreover, we shall also use the momentum dependent polarization function to estimate the permittivity of the medium which will be used later to estimate the heavy quark potential. We note that such a calculation for the Debye mass for the photons was carried out in Refs.\cite{Alexandre:2000jc,Fukushima:2015wck,Bandyopadhyay:2016fyd}. In the following, we briefly outline the calculations as in Ref.\cite{Bandyopadhyay:2016fyd} taking care of the appropriate color factors.
Consider that a charged particle of charge $q_f$ and mass $m_f$ is moving in a constant magnetic field ($ B $) which is acting along the z-direction, i.e., $\vec{B}=B \hat{z}$. In this scenario the propagator of the charged particle is a function of both the transverse and longitudinal component of the momentum ( with respect to $ B $), which is due to the breaking of Lorentz invariance.
 
The fermion propagator in coordinate space is written as~\cite{Schwinger:1951nm}
\begin{eqnarray}
S(y,y')&=&\Phi(y,y')\int \frac{d^4 k}{(2 \pi)^4} e^{-i {k}\cdot ({y}-y')}S(K),
\label{eqn01} 
\end{eqnarray}
where $\Phi(y,y')$ is the phase factor and can be gauged away. In Eq.~(\ref{eqn01}), 
$S(K)$ is the Fourier transform of the fermion propagator. It is convenient to write the propagator in the Landau level representation. Proceeding in a similar way as done in Ref.~\cite{Chodos:1990vv} for Euclidean space, the resulting propagator can be written as
\begin{equation} 
S(K)=-i e^{-\frac{K_{\perp}^2}{q_fB}}\sum_{l=0}^{\infty}\frac{(-1)^lD_l(q_fB,K)}{K_{\parallel}^2+m_{f}^2+2l|q_fB|},
\label{eqn0}
\end{equation}
where the sum is over all the Landau levels ($l=0,1,2,$....), and 
\begin{eqnarray}
D_l(q_fB,K)&=&(m_f-\slashed{K}_{\parallel})\bigg((1+i\gamma_1 \gamma_2)L_l\bigg(\frac{2K_{\perp}^2}{q_fB}\bigg)
-(1-i\gamma_1 \gamma_2)L_{l-1}\bigg(\frac{2K_{\perp}^2}{q_fB}\bigg) \nonumber\\
&+&4\slashed{K}_{\perp}L_{l-1}^1\bigg(\frac{2K_{\perp}^2}{q_fB}\bigg) \bigg),
\end{eqnarray}

where $K_{\parallel}=(K_4, K_3)$, $K_{\perp}=(K_1, K_2)$ are the 4-momentum component parallel and perpendicular to the magnetic field with $K_4=-i K_0$ and $\gamma_1, \gamma_2$ are the Dirac matrices.\\
The energy spectrum of a charged fermion %of charge, $q_f$ and mass, $ m_f $ 
in the presence of magnetic field can be written as \cite{beres}
\begin{equation} 
E_l^2(K_3)=K_3^2+m_f^2+2l|q_f B|,
\label{spectrum}
\end{equation}
which can be obtained from Eq.~(\ref{eqn0}) by equating the denominator of the propagator to zero. From Eq.~(\ref{spectrum}), it is clear that the energy of fermion is discrete in the direction  perpendicular to $ B $ and continuous in the direction parallel to the magnetic field. The discretization of energy levels is known as Landau levels.

In the presence of a very strong magnetic field, i.e., $q_f B \gg T^2$ the higher Landau levels ($l\gg 1$) are at infinity as compared to LLL \cite{Bandyopadhyay:2016fyd,Calucci:1993fi} due to which the LLL  dominates and the dimensional reduction takes place.
For LLL approximation 
one can use the relations $L_n(x)=L_{n}^{0}(x)$ and $L_{-1}^{\alpha}(x)=0$
(by definition). In this approximation, the fermion propagator Eq.~(\ref{eqn0}) becomes
\begin{equation}
S(K)=ie^{-\frac{K_{\perp}^2}{q_fB}}\bigg(\frac{\slashed{K}_{\parallel}-m_f}{K_{\parallel}^2+m_f^2}\bigg)(1+i \gamma_1 \gamma_2),
\label{eqn1}
\end{equation}
where the factor $(1+i\gamma_1 \gamma_2)$ is the projection operator which appears because the spin of the fermion in the LLL is polarized parallel to the magnetic field hence signifies the spin-polarized nature of LLL \cite{Miransky:2015ava}. The spin orientation is parallel/antiparallel to the magnetic field direction for the positively/negatively charged fermion. The form of the propagator in Eq.~(\ref{eqn1}) clearly demonstrates the dimensionally reduced character $ (1+1) $ of the LLL. It is evident that the dimensional reduction restricts the motion of charged particles perpendicular to the magnetic field.  

\begin{figure}
\includegraphics[scale=0.75]{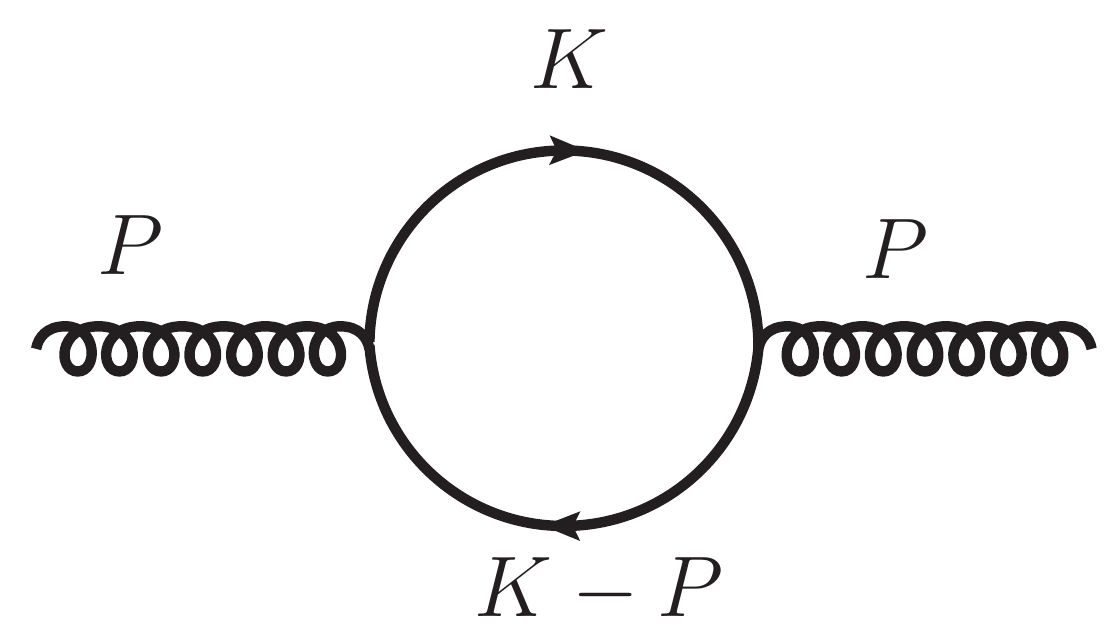}
\caption{	Gluon self-energy in the presence of strong magnetic field}.
\label{fig1}
\end{figure}

For evaluating the gluon polarization tensor and Debye screening mass we use the following identities
\begin{equation}\nonumber 
b_{\parallel}^{\mu}=(b_4,0,0,b_3); \hspace{0.5cm} a_{\parallel}^{2}=a_{4}^{2}+a_{3}^{2}; \hspace{0.5cm} b_{\perp}^{\mu}=(0,b_1,b_2,0),
\end{equation}
\begin{equation}\nonumber 
(a \cdot b)_{\parallel}=a_{4}b_{4}+a_{3}b_{3}; \hspace{0.5cm}
(a \cdot b)_{\perp}=a_{1}b_{1}+a_{2}b_{2},
\end{equation}
where $ \parallel $ and $ \perp $ are the parallel and perpendicular components.
In a background magnetic field, the only contribution to the gluon self-energy comes from the quark loop as shown in  Fig.~\ref{fig1}. As gluons do not interact with the magnetic field so their contribution remains the same as that at $ B=0 $.
%%%%%%%%%%%%%%%%%%%%%%%%%%%%%%%%%%%%%%%%%%%%%%%%%%%%%%%%%%%%%%
%%%%%%%%%%%%%%%%%%%%%%%%%%%%%%%%%%%%%%%%%

Therefore, the quarks contribution to the gluon self-energy in the presence of magnetic field, $ \delta\Pi_{\mu \nu}(P,B) $ can be written as
\begin{eqnarray} 
\delta\Pi_{\mu \nu}(P,B) = \int \frac{d^4 K}{(2 \pi)^4} Tr[(g t^a\gamma_{\mu})S(K)(g t^b\gamma_{\nu})S(K-P)],
\label{eqn2}
\end{eqnarray}
where S(K) is fermion propagator as defined in Eq.~(\ref{eqn1}). As the dimensional reduction separate the parallel and perpendicular component of the momentum in the propagator. So the self-energy can also be separated into components.
The perpendicular part of the self energy in Eq.~(\ref{eqn2}) is written as

\begin{eqnarray} 
I_{\perp}&=&\frac{1}{(2 \pi)^2}\int_{-\infty}^{\infty}\int_{-\infty}^{\infty} dK_x dK_y \exp[\frac{1}{q_f B}(-K_x^2-(K_x-P_x)^2)] \exp[\frac{1}{q_f B}(-K_y^2-(K_y-P_y)^2)]\nonumber\\
&=&\sum
_{f}\frac{\pi q_f B}{2 (2\pi)^2}\exp\bigg(\frac{-P_{\perp}^2}{2 q_f B}\bigg).
\label{Iperp} 
\end{eqnarray}
Therefore, Eq.~(\ref{eqn2}) becomes
\begin{equation} 
\delta\Pi_{\mu \nu}(P,B)=g^2 Tr(t^a t^b)I_{\perp} \int \frac{d^2 K_{\parallel}}{(2 \pi)^2}\frac{Tr[\gamma_{\mu}(\slashed{K}_{\parallel}-m_f)(1+i \gamma_1 \gamma_2)\gamma_{\nu}((\slashed{K}-\slashed{P})_{\parallel}-m_f)(1+i \gamma_1 \gamma_2)]}{(K_{\parallel}^2+m_f^2)((K-P)_{\parallel}^2+m_f^2)}.
\end{equation}

In the imaginary time formalism, the temporal component of self energy becomes
\begin{equation}
\delta\Pi_{44}(P,B)= g^2 Tr(t^a t^b)I_{\perp}\int \frac{d K_{3}}{(2 \pi)}T\sum_{n}\frac{8 \omega_n^2-8K_3^2-8m_f^2-8\omega_n \omega+8 K_3P_3  }{(\omega_n^2+K_3^2+m_f^2)((\omega-\omega_n)^2+(K_3-P_3)^2+m_f^2)},
\label{eqnmfPi44}
\end{equation}
where $\omega_n=(2n+1)\pi T$ is fermionic Matsubara frequency  and $\omega$ is bosonic Matsubara frequency. For massless fermion ($m_f=0$) we get
\begin{equation}
\delta \Pi_{44}(B)= 8g^2 Tr(t^a t^b)I_{\perp}\int \frac{d K_{3}}{(2 \pi)}T\sum_{n}\bigg(\tilde{\Delta}(K)-(2K_3^2+\omega_n\omega-K_3P_3)\tilde{\Delta}(K)\tilde{\Delta}(K-P)\bigg),
\label{eqnPi44}
\end{equation}
where $\tilde{\Delta}(K)=\frac{1}{\omega_n^2+K_3^2}$, $\tilde{\Delta}(K-P)=\frac{1}{(\omega - \omega_n)^2+(K_3-P_3)^2}$. In order to calculate the Debye screening mass one can take the static limit of temporal component of self-energy, $ \Pi_{44} $ {\em viz.} $m_D^2=-\Pi_{44}$($\omega \rightarrow 0$, $\vec{P}=0$). After taking the static limit Eq.~(\ref{eqnPi44}) becomes
\begin{eqnarray}
\delta \Pi_{44}(B)|_{(\omega \rightarrow 0,\vec{P}=0)}&=& 8g^2 Tr(t^a t^b)I_{\perp}\int \frac{d K_{3}}{(2 \pi)}T\sum_{n}\bigg(\tilde{\Delta}(K)-2K_3^2(\tilde{\Delta}(K))^2\bigg)
\label{eqn44}
\end{eqnarray}

 After doing the sum over Matsubara frequency mode in  Eq.~(\ref{eqn44}) as in Refs. \cite{Bellac:2011kqa,Bandyopadhyay:2016fyd}, one can get

\begin{equation}
\delta m_D^2= 4 g^2 \beta I_{\perp} \int \frac{d K_{3}}{(2 \pi)} \tilde{f}(E)(1-\tilde{f}(E)).
\end{equation}
By using Eq.~(\ref{Iperp}), the Debye screening mass in the presence of the magnetic field becomes
\begin{equation}
\delta m_D^2=\sum_{f}\frac{|{q_f}|B g^2}{ \pi T}\int_{0}^{\infty}\frac{d K_3}{2\pi}\tilde{f}(E)(1-\tilde{f}(E)).
\label{eqnn14}
\end{equation}

By taking the gluonic contribution into account the total Debye mass becomes
\begin{equation}
m_D^2 = \frac{4 \pi \alpha_s(T) T^2 N_c}{3}+\sum_{f}\frac{ |{q_f}|B g^2}{ \pi T}\int_{0}^{\infty}\frac{dK_3}{2\pi}\tilde{f}(E)(1-\tilde{f}(E)).
\label{eqn3}
\end{equation}
\begin{figure}[tbh]
	\subfigure{
		\hspace{-0mm}\includegraphics[width=8cm]{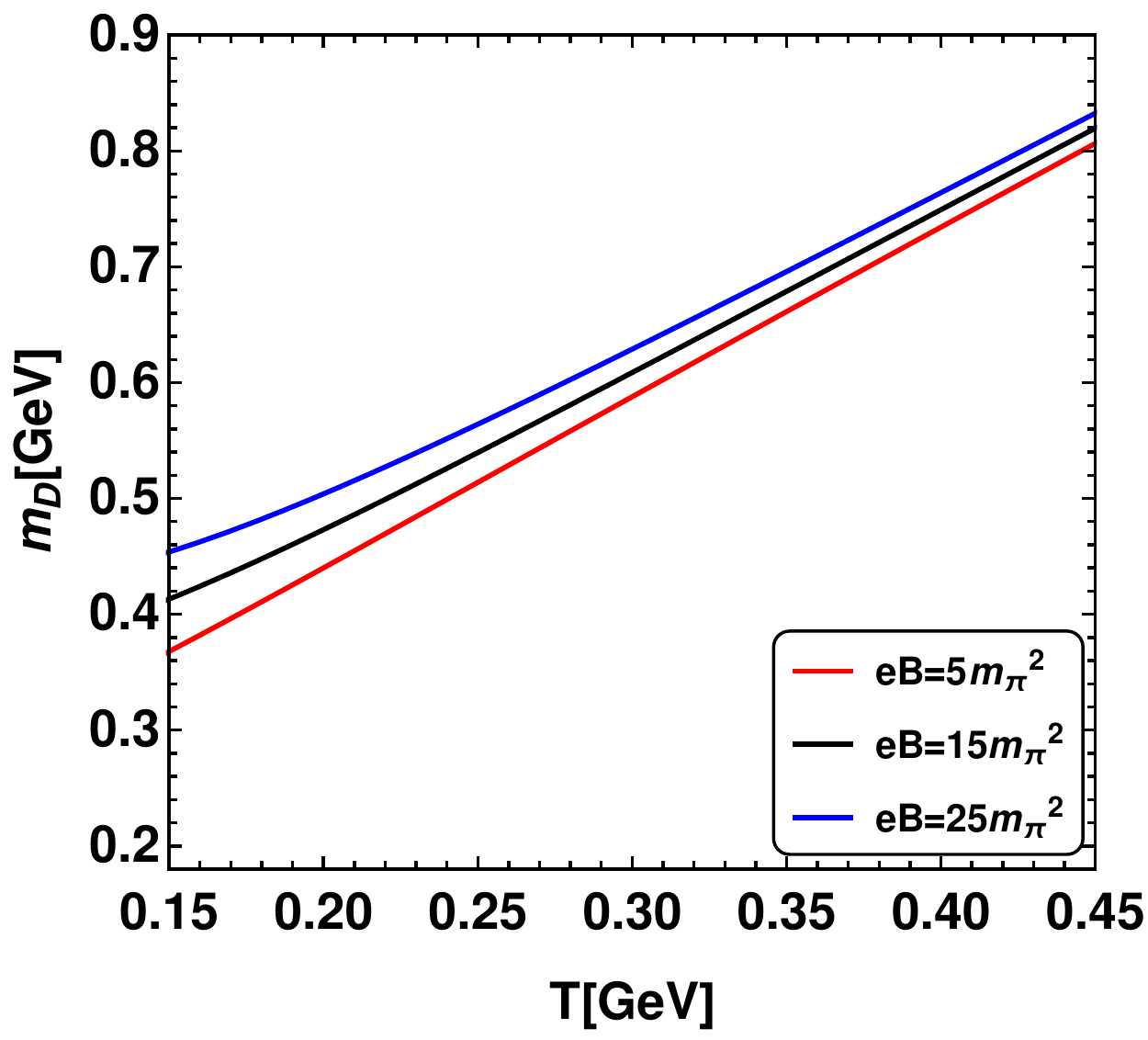}} 
	\subfigure{
		\includegraphics[width=7.7cm]{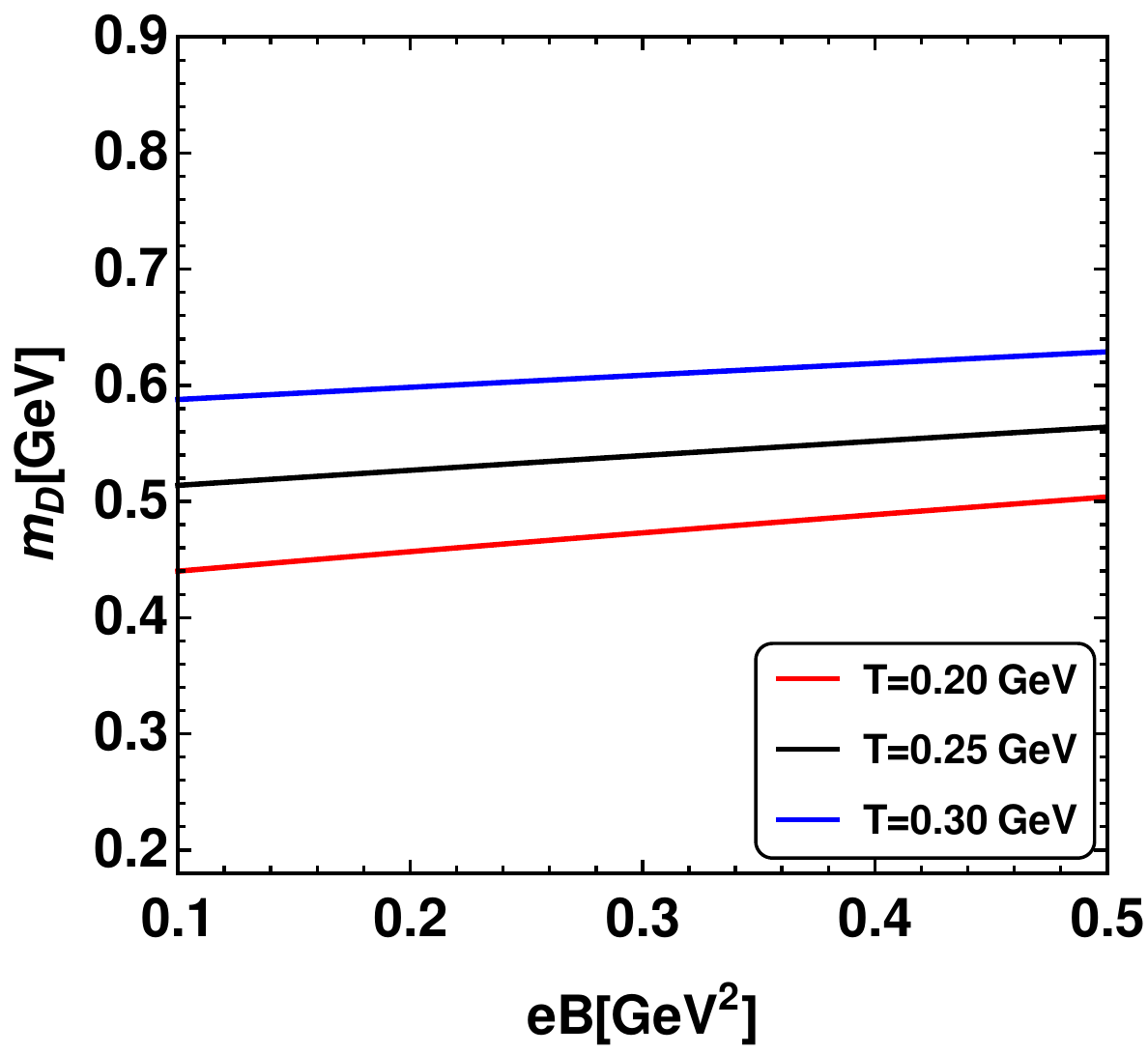}}
	\caption{
		Left panel: Variation of the Debye screening mass ($ m_D $) with temperature for different values of magnetic field ($ eB $). Right panel: Variation of $ m_D $ with magnetic field for different values of $ T $.}

	\label{mD}
\end{figure}
For massless fermions, $\int_{0}^{\infty}dK_3\tilde{f}(E)(1-\tilde{f}(E)=T/2$. %\frac{T}{2}$.
Therefore, Eq. (\ref{eqn3})
reduces to
\begin{equation}
m_D^2 = \frac{4 \pi \alpha_s(T) T^2 N_c}{3}+\sum_{f}\frac{ |{q_f}|B \alpha_s(T)}{ \pi}.
\label{mDB}
\end{equation}
where $ \alpha_s(T) $ is the one loop coupling constant %~\cite{Bannur:2006js,Zhu:2009zzi} {\color{red}change cite for one loop}. 
and which can be written as
\begin{eqnarray}
\alpha_{s}(T)=\frac{g_s^{2}(T)}{4 \pi}=\frac{6 \pi}{\left(33-2 N_{f}\right)\ln \left(\frac{2\pi T}{\Lambda_{\overline{\rm MS}}}\right)}.
\end{eqnarray}
Here we use $ N_f=3 $ and $\Lambda_{\overline{\rm MS}} = 0.176$ GeV~\cite{Haque:2014rua}.

In the left panel, Fig.~\ref{mD} shows the variation of Debye screening mass for $ m_f=0 $ with $ T $ for different values of magnetic field, i.e., $ eB=5~m_\pi^2 $, $ 15~m_\pi^2 $ and $ 25~m_\pi^2 $, respectively. In the right panel, the figure shows the variation of Debye screening mass with $ B $ for different values of temperature, i.e., $ T=0.20 $, $ 0.25 $ and $ 0.30 $ GeV, respectively.
Here the Debye screening mass Eq.~(\ref{mDB}) depends on the two scales, i.e., $ T $ and $ B $. 
We find that the Debye screening mass increases with the increase in temperature and magnetic field in a magnetized hot QCD medium. Indeed, the effect of magnetic field is stronger at a lower temperature and becomes weaker at a higher temperature. Therefore, we can say that there is a finite amount of Debye screening with $ T $ and $ B $ for $ m_f=0 $. The observed increase of screening mass as a function of magnetic field is in qualitative agreement with lattice QCD computation~ \cite{Bonati:2017uvz}.

%%%%%%%%%%%%%%%%%%%%%%%%%%%%%%%%%%%%%%%%%%%%%%%%%%%%%%%%%%% 
\section{In-medium heavy quark potential in magnetic field}
\label{II}
In this section we study the modification of the
Cornell potential in the presence of a hot medium endowed also with a magnetic field. Before that let us first discuss the heavy quark-antiquark potential for a vanishing magnetic field.
Consider that a test charged particle is placed at the origin and an auxilar vector field due to this is written as $\vec{E}=qr^{b-1}\hat{r}$, where $b$ is a parameter that  can have any value. The corresponding potential can be defined by using the relation -$\vec{\nabla}V(r)=\vec{E}(r)$. Therefore, the Gauss law can be defined as~\cite{Burnier:2015nsa,Dixit:1989vq} 
\begin{equation}
\vec{\nabla}.\bigg(\frac{\vec{E}}{r^{b+1}}\bigg)=4\pi q \delta(r).
\label{Gausseq}
\end{equation}

As per the Debye-H\"{u}ckle framework, the medium gets polarized and leads to
a change of the source term on the right hand side of Eq. (\ref{Gausseq}) from $ \delta(r) $ to $ \delta(r)+ \langle \rho(r) \rangle$.
Here $\langle \rho(r) \rangle$
is the induced charged density and can be written in the Boltzmann's distribution as the difference of the particle and antiparticle charge deviation. 
At high temperatures and weak potential the charge density can be approximated as~
$ \langle \rho(r) \rangle=-2q\beta n_0V(r) $,
which is equivalent to the linear response approximation. Here $ \beta=1/T $ and $ n_0 $ is the charge density in the absence of test charge.
After substituting the above relation into Eq.~(\ref{Gausseq}) and by
choosing the appropriate values of $ b $ and $ q $, one can obtain the Coulombic and string part of the potential.
For $b=-1, q={\alpha} (={\alpha}_sC_F=\frac{g_s^2C_F}{4\pi}; C_F=4/3)$,  Eq.~(\ref{Gausseq}) reduces to the Coulomb potential and 
and for $b=1, q=\sigma$, one can obtain the linearly rising (string part) potential, where $\sigma$ is the string tension and ${\alpha}_s$ is the QCD running coupling.
The detailed calculation of the potential for finite T
in the Debye-H\"{u}ckel approach can be found in Ref.~\cite{Burnier:2015nsa}.\\

In a quantum field theory framework, the medium effects, i.e., the effect of finite temperature and magnetic field can be incorporated by modifying the vacuum potential with dielectric permittivity as in Ref.~\cite{Thakur:2013nia}. The in-medium permittivity $ \epsilon(\vec{p},m_D) $ can be written as
\begin{equation}
\epsilon^{-1}(\vec{p},m_D)=-\lim_{\omega \to 0} {p^2} D^{00}(\omega, p)~.
\label{ephs}
\end{equation}
where $D^{00}$ is the gluon propagator calculated from the gluon self energy. This will have contributions from the gluonic and fermionic loop. We evaluate each of the contributions in the presence of magnetic field in the following way. 

\subsection{Gluon contribution}
As the gluonic contribution does not depend upon the magnetic field, the longitudinal part of the gluon self-energy can be written as ~\cite{Chen:2017zwe} 
\begin{equation}
\delta\Pi_L(\omega,p)_g=m_{Dg}^2\bigg[1-\frac{\omega}{2 p}\ln\bigg(\frac{\omega+p}{\omega-p}\bigg)+i\pi\frac{\omega}{2 p}\theta(p^2-\omega^2)\bigg]
\end{equation}
where $m_{Dg}^2=\frac{1}{3} g^2 T^2N_c$ and $\theta(p^2-\omega^2)$ is the step function. In the above equation, the imaginary part is related to the Landau damping,  which corresponds to the emission and absorption of particle in the medium.
\subsection{Fermion contribution} 
From Eq.~(\ref{eqnmfPi44}), we can get $\delta\Pi_L(\omega,p)\equiv\delta\Pi_{44}(\omega,p)$ after doing sum over fermionic Matsubara frequencies. There are two types of frequency sum in Eq.~(\ref{eqnmfPi44}) which we evaluate here in detail. The first type 
is given by
\begin{eqnarray}
T\sum_n \frac{1}{(\omega-\omega_n)^2+E_Q^2}&=&T\sum_n\frac{1}{2 E_Q}\bigg[\frac{1}{E_Q+i(\omega-\omega_n)}-\frac{1}{E_Q-i(\omega-\omega_n)}\bigg]\nonumber\\
&=&\frac{1}{4 E_K}\bigg[\tanh\bigg(\frac{E_K-\omega}{2 T}\bigg)+\tanh\bigg(\frac{E_K+\omega}{2 T}\bigg)\bigg]
\label{sum1}
\end{eqnarray}
and the second type is 
\begin{eqnarray}
T\sum_n \frac{1}{((\omega-\omega_n)^2+E_Q^2)(\omega_n^2+E_K^2)}&=&T\sum_n\frac{1}{4 E_K E_Q}\bigg[\frac{1}{(E_K+i\omega_n)(E_Q+i(\omega-\omega_n))}\nonumber\\
&+&\frac{1}{(E_K-i\omega_n)(E_Q+i(\omega-\omega_n))}\nonumber\\
&+&\frac{1}{(E_K+i\omega_n)(E_Q-i(\omega-\omega_n))}\nonumber\\
&+&\frac{1}{(E_K-i\omega_n)(E_Q-i(\omega-\omega_n))}\bigg].
\end{eqnarray}
Using the summation over the discrete Matsubara frequencies ($\omega_n=(2 n+1)\pi T$) above equation becomes
\begin{eqnarray}
T\sum_n \frac{1}{((\omega-\omega_n)^2+E_Q^2)(\omega_n^2+E_K^2)}&=&\frac{1}{4 E_Q E_K}\bigg[\frac{\tilde{f}(E_K)-\tilde{f}(E_Q-\omega)}{\omega+E_K-E_Q}+\frac{\tilde{f}(E_K)-\tilde{f}(E_Q+\omega)}{E_K-E_Q-\omega}\nonumber\\
&+&\frac{1-\tilde{f}(E_K)-\tilde{f}(E_Q-\omega)}{E_K+E_Q-\omega}+\frac{1-\tilde{f}(E_K)-\tilde{f}(E_Q+\omega)}{E_K+E_Q+\omega}\bigg].
\label{sum2}
\end{eqnarray}
In the above, the expressions are written after doing an analytic continuation to Minkowski space. The imaginary part of the longitudinal component of the self-energy for fermion loop ($\Im\delta\Pi_{44}$) can be calculated by using the identity
\begin{equation}
\Im \delta\Pi_L(\omega,p)_f=\frac{1}{2i}\lim\limits_{\eta\rightarrow0}\bigg[\delta\Pi_L(\omega+i\eta,p)-\delta\Pi_L(\omega-i\eta,p)\bigg]
\label{img}
\end{equation}
along with the expression
\begin{equation}
\frac{1}{2i}\bigg(\frac{1}{\omega+\sum_{j}E_j+i\eta}-\frac{1}{\omega+\sum_{j}E_j-i\eta}\bigg)=-\pi\delta(\omega+\sum_j E_j).
\end{equation}
Using Eqs.~(\ref{sum1}) and (\ref{sum2}) in Eq.~(\ref{eqnmfPi44}) the imaginary part of $\delta \Pi_L$  reduces to
\begin{eqnarray}
\delta\Im{\Pi_L}(\omega,p)_f&=&-\pi 4 g^2 I_{\perp}\int \frac{dK_3}{2 \pi}\bigg(\frac{-2K_3^2+K_3P_3+2m_f^2}{4 E_Q E_K}\bigg)\bigg[(-\tilde{f}(E_K)+\tilde{f}(E_Q-\omega))\nonumber \\
&\times&\delta(\omega+E_K-E_Q)+(\tilde{f}(E_K)-\tilde{f}(E_Q+\omega))\delta(\omega-E_K+E_Q)+(1-\tilde{f}(E_K)\nonumber \\
&-&\tilde{f}(E_Q-\omega))\delta(E_K+E_Q-\omega)+(1-\tilde{f}(E_K)-\tilde{f}(E_Q+\omega))\delta(E_K+E_Q+\omega)\bigg].
\label{img1}
\end{eqnarray}
 
In the limit $\omega\rightarrow0$,
the third and fourth terms will vanish. Therefore, the above equation becomes
\begin{equation}
\delta\Im \Pi_L(\omega,p)_f|_{\omega\rightarrow0}=\pi  \omega 4 g^2\int\frac{dK_3}{2 \pi}\bigg(\frac{-2K_3^2+K_3P_3+2m_f^2}{4 E_Q E_K}\bigg) \frac{\partial f(E_Q)}{\partial E_Q}\delta(E_K-E_Q).\,\,\,\,\,\,
\end{equation}

Here, the integral can be solved by using the properties of the Dirac delta function as
\begin{equation}
\delta(f(x))=\sum_n\frac{\delta(x-x_n)}{|\frac{\partial f(x)}{\partial x}|_{x=x_n}}
\end{equation}
where $x_n$ are the zeros of the function $ f(x) $. In Eq.~(\ref{img1}) delta functions give zeros as
\begin{equation}
K_{30}=\frac{4 P_3 (P_3^2-\omega^2)\pm \sqrt{16P_3^2(P_3^2-\omega^2)^2-16(P_3^2-\omega^2)((P_3^2-\omega^2)^2-4m^2\omega^2)}}{8 (P_3^2-\omega^2)}
\end{equation}

In the limit $\omega \rightarrow 0$, we have only one zero of the function, $K_{30}=P_{3}/2$, so that
\begin{equation}
\Im \delta\Pi_{L}(p)=\frac{ \omega \beta g^2 m_f^2 I_{\perp}}{P_3 E} (\tilde{f}(E)-\tilde{f}^2(E)) 
\end{equation}
where $E=\sqrt{m_f^2+\frac{P_3^2}{4}}$. For calculating the fermion loop contribution to the imaginary part of the gluon propagator, we use the spectral function approach in which the propagator is related to real and imaginary parts of the self-energy, it can be written as ~\cite{Weldon:1990iw} 
\begin{equation}
\Im{D^{\mu \nu}}(\omega,P)=-\pi(1+e^{-\beta \omega})\xi^{\mu \nu}.
\end{equation}
Here, we focus only on $D_{L}\equiv D^{00}$, for this purpose we write $\xi^{00}$ as
\begin{equation}
\xi^{00}(\omega,P)=\frac{1}{\pi}\frac{e^{\beta \omega}}{e^{\beta \omega}-1}\rho_L(\omega,P)
\end{equation}	
where $\rho_L$ is the longitudinal part of spectral function which describes the quasiparticles with finite width. In Breight-Wigner form it is written as
\begin{equation}
\rho_L(\omega,P)=\frac{\Im{\Pi_L}}{(P^2-\Re{\Pi_L})^2+\Im{\Pi_L}^2}.
\end{equation}
Taking fermion and gluon contributions into account, real and imaginary parts of self-energy can be written as 
\begin{equation}
\Re \Pi_L(\omega,p)=\Re{\Pi_L(\omega,p)_g}+\Re{\Pi_L(\omega,p)_f},
\end{equation}
\begin{equation}
\Im \Pi_L(\omega,p)=\Im{\Pi_L(\omega,p)_g}+\Im{\Pi_L(\omega,p)_f}.
\end{equation}	
Here, $\Re{\Pi_L}$ is the real part of the longitudinal component of self-energy, which is the square of the Debye screening mass ($m_D^2$) in the limit $\omega\rightarrow0$, (Eq.~(\ref{mDB})).

Therefore, using the expression for real and imaginary parts of the self-energy in the limit $\omega\rightarrow0$, we have calculated the gluon propagator ($ D^{00} $) in terms of real and imaginary parts as
\begin{equation}
D^{00}(p)=\Re D^{00}(p)+\Im D^{00}(p).
\end{equation}

\begin{equation}
D^{00}(p)=\frac{-1}{p^2+m_D^2}+\frac{i\pi T m_{Dg}^2}{p(p^2+m_D^2)^2}-\frac{i|q_f B| m_f^2\alpha_s}{2 (p^2+m_D^2)^2 P_3 E \cosh^2(\frac{\beta E}{2})}.
\end{equation}
Note that due to the specific direction of the magnetic field the contribution of the fermion field loop makes the propagator anisotropic. However, in the limit of vanishing light quark mass the propagator remains still isotropic with the effect of magnetic field showing only in the Debye mass. Thus, the dielectric permittivity can be calculated from Eq.~(\ref{ephs}) and
in the limit of massless fermions, it becomes
\begin{equation}
\epsilon^{-1}(\vec{p},m_D)=\frac{p^2}{p^2+m_D^2}-i\pi T\frac{p    m_{Dg}^2}{(p^2+m_D^2)^2}.
\label{eqn13}
\end{equation}
%Few comments at this point may be in order. 
From the above equation it is clear that the permittivity is isotropic in the limit of vanishing light quark masses. Further, for an effective description of quarkonium in terms of a potential at finite temperature, the mass of heavy quark , $m_Q $ should be much larger than  $ \varLambda_{QCD} $ as well as  $m_Q \gg T$. For the magnetic field considered here, $ m_Q $ is still the largest scale ($m_Q \gg \sqrt{eB} $) i.e., the ratio, $ m_Q^2/eB\simeq 3-15 $ for the range of magnetic field $ eB=5-25~ m_\pi^2 $, so that the interaction between quark and antiquark can be described by a quantum mechanical potential. Therefore, our approximation to take the static heavy quark potential for heavy quarkonia should be valid here. This has been attempted here to obtain the effects of the medium on the vacuum potential by correcting both the short and long distance part by a dielectric function encoding the effects of the magnetized deconfined medium. 

The Coulomb part of the potential in Fourier space in the presence of medium can be written as~\cite{JDJackson}
\begin{equation}
k^2 V_C(\vec{p})=4\pi \frac{{\alpha}}{\epsilon(\vec{p},m_D)}
\label{eqn12}
\end{equation}
The Fourier transformation of Eq.~(\ref{eqn12}) in coordinate space gives
\begin{equation}
-\nabla^2V_C(r)+m_D^2V_C(r)={\alpha}(4\pi \delta(r)-iTm_{Dg}^2 h(m_D r))
\label{eqn14}
\end{equation}
where $ h(y)=2\int_{0}^{\infty}dx\frac{x}{(x^2+1)}  \frac{\sin(yx)}{yx} $.

In a similar manner, one can calculate the string part of the potential and get the differential equation for $ V_S(r) $ as~\cite{Burnier:2015nsa} %(Eq.~(\ref{eqn11})) as
\begin{equation}
-\frac{1}{r^2}\frac{d^2V_S(r)}{dr^2}+\mu^4 V_S(r)=\sigma(4\pi \delta(r)-i Tm_{Dg}^2h(m_D r)),
\label{eqn15}
\end{equation}
where $\mu=(\frac{m_{Dg}^2 \sigma}{\alpha})^{1/4}$.
 
After using the boundary conditions  $m_D\rightarrow 0$, $\Re V_C(r)=-\frac{\alpha}{r}$ and  $r\rightarrow 0$, $\Im V_C(r)=0$~\cite{Digal:2005ht}, one can obtain both the real and imaginary part of the Coulomb potential as
\begin{equation}
\Re V_C(r,T,B)=-{\alpha}\frac{e^{-m_D r}}{r}-{\alpha} m_D,
\end{equation}
and 
\begin{equation}
\Im V_C(r,T,B)=-2{\alpha}Tg(m_Dr), 
\end{equation}

where $g(y)=\int_{0}^{\infty}dx\frac{x}{(x^2+1)^2}\bigg(  1-\frac{\sin(yx)}{yx} \bigg) $. Thus, the magnetic field dependence of the potential arises from the field dependent Debye mass.

Similarly, for the string part of the potential we use the following boundary conditions $ \mu \rightarrow 0, \Re V_S(r)=\sigma r$, $ r\rightarrow 0, \Im V_S(r)=0 $ and $  r\rightarrow\infty, \frac{d \Im V_S(r)}{dr}=0$. After using the boundary conditions we get both the real and imaginary parts of string potential as 
\begin{equation}
\Re V_S(r,T,B)=-\frac{\Gamma(\frac{1}{4})}{2^{\frac{3}{4}}\sqrt{\pi}}\frac{\sigma}{\mu}D_{-\frac{1}{2}}(\sqrt{2}\mu r)+\frac{\Gamma(\frac{1}{4})}{2\Gamma(\frac{3}{4})}\frac{\sigma}{\mu},
%\label{eqn16}
\end{equation}
where $D_{\nu}(x)$ is the parabolic cylinder function, and
\begin{equation}
\Im V_S(r,T,B)=-\frac{\sigma m_{Dg}^2T}{\mu}\phi(\mu r),
\end{equation}
where
	\begin{eqnarray}
	\phi(\mu r)&=& D_{-\frac{1}{2}}(\sqrt{2}\mu r)\int_{0}^{r}dx \Re D_{-\frac{1}{2}}(i \sqrt{2}\mu x) x^2 g(m_D x)
	+\Re D_{-\frac{1}{2}}(i \sqrt{2}\mu r)\int_{r}^{\infty}dx D_{-\frac{1}{2}}(\sqrt{2}\mu x) x^2 g(m_D x)\nonumber\\
	&-&D_{-\frac{1}{2}}(0)\int_{0}^{\infty}dx  D_{-\frac{1}{2}}(\sqrt{2}\mu x) x^2 g(m_D x),\nonumber
	\end{eqnarray}
The total real part of the potential after combining both the Coulombic and string parts in a magnetized medium can be written as
\begin{eqnarray}
\Re V(r,T,B)&=&\Re V_C(r,T,B)+\Re V_S(r,T,B)\nonumber\\
&=&-{\alpha}\frac{e^{-m_D r}}{r}-{\alpha} m_D -\frac{\Gamma(\frac{1}{4})}{2^{\frac{3}{4}}\sqrt{\pi}} \frac{\sigma}{\mu}D_{-\frac{1}{2}}(\sqrt{2}\mu r)+\frac{\Gamma(\frac{1}{4})}{2\Gamma(\frac{3}{4})}\frac{\sigma}{\mu}.
\end{eqnarray}
Similarly, the total imaginary part of the potential becomes

\begin{eqnarray}
\Im V(r,T,B)&=&\Im V_C(r,T,B)+\Im V_S(r,T,B)\nonumber\\
&=&-2{\alpha} Tg(m_Dr)
-\frac{\sigma m_{Dg}^2T}{\mu}\phi(\mu r),
\label{imagpt}
\end{eqnarray}
In the framework of Debye-H\"{u}ckel theory,
the Coulomb and string terms were modified with different screening scales $ m_D $ and $ \mu $, respectively, to obtain
the heavy quark potential. On the other hand in Ref.~\cite{Hasan:2017fmf} the authors have used the same screening scale, $ m_D $ for both the
Coulombic and linear terms. It is interesting to see the effects of
different scales for the Coulombic (perturbative) and the linear (nonperturbative) terms.
%%%%%%%%%%%%%%%%%%%%%%%%%%%%%%%%%%%%%%Figure1%%%%%%%%%%%%%%%
\begin{figure}[tbh]
	\subfigure{
		\hspace{-0mm}\includegraphics[width=8cm]{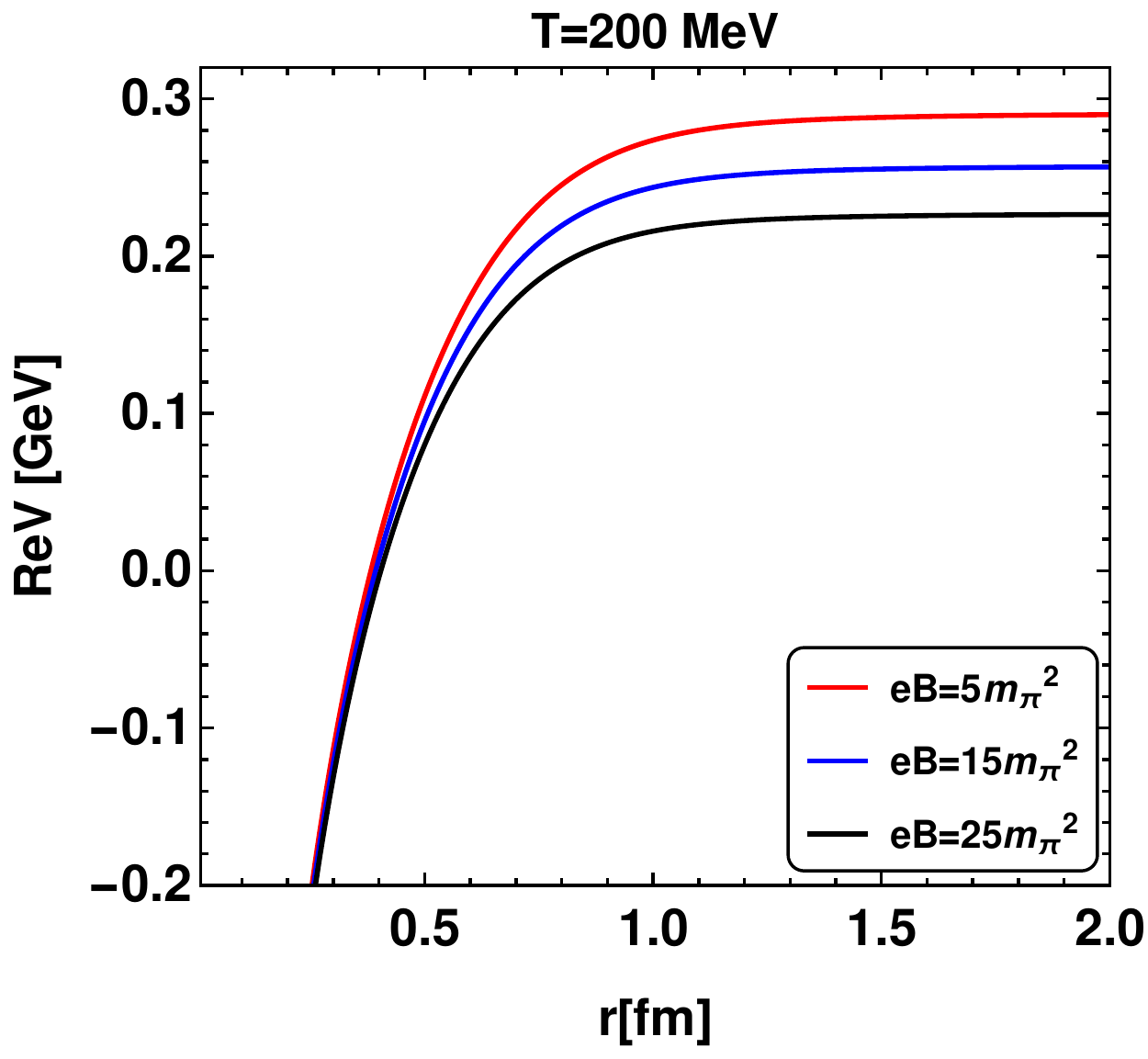}} 
	\subfigure{
		\includegraphics[width=8cm]{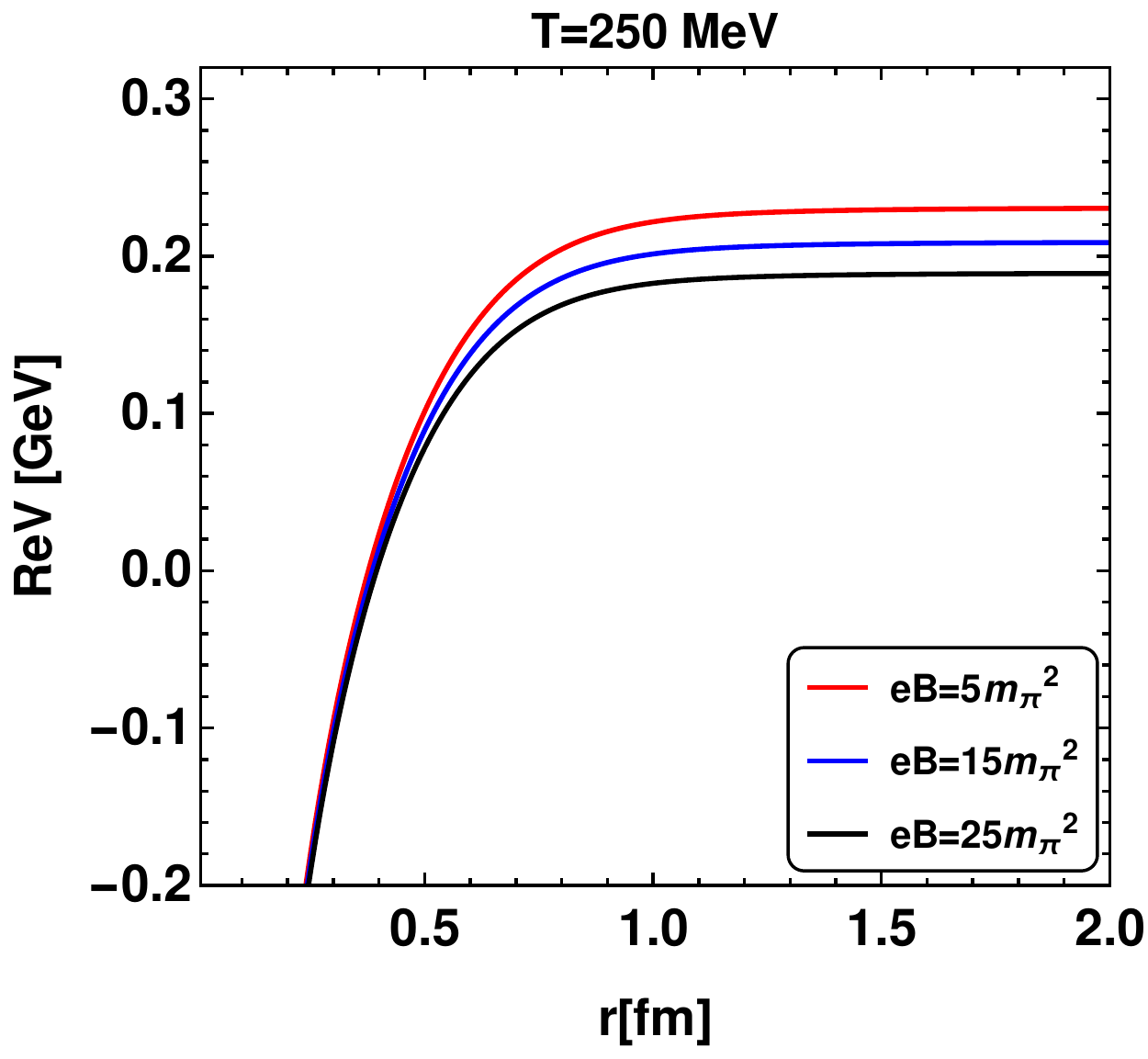}}
	\caption{
	Variation of the real part of potential with separation distance $ r $ between $Q\bar{Q}$ for various values of magnetic field %, i.e., $~eB= 15 m_{\pi}^2,~ 30 m_{\pi}^2, ~45 m_{\pi}^2 $ 
	at $T=200$ MeV (left) and $T=250$ MeV (right).}
	%\label{pot_re_theta}
	\label{fig2}
\end{figure}
%%%%%%%%%%%%%%%%%%%%%%%%%%%%%%%%%%%%%%%%%%%%%%%%%%%%%%%%%%%
%%%%%%%%%%%%%%%%%%%%%%%%%%%%%%%%%%%%%%%%%%%%%%%%%%%%%%%

Figure~\ref{fig2} shows the variation of the real part of the potential  with the separation distance ($ r $) between the Q$\bar{Q}$ pair for different values
of magnetic field ($~eB= 5 m_{\pi}^2,~ 15 m_{\pi}^2, ~25 m_{\pi}^2 $~) at $ T= 200 $ MeV (left) and $ T= 250 $ MeV (right). Here, we use
the value of the string tension, $ \sigma = 0.174  $ GeV$^2$ from Ref.~\cite{Burnier:2015nsa} . From the figure we find that the screening increases with the increase in magnetic field. 
The screening is more at a higher temperature ($ T= 250 $ MeV) as compared to lower temperature ($ T= 200 $ MeV) because at higher temperatures
the quarkonium state is loosely bound as compared to the lower
temperatures and gets easily dissociated. Alternatively, we can say that with the increase in temperature the gluonic contribution becomes more which results in more screening.

%%%%%%%%%%%%%%%%%%%%%%%%%%%%%Figure2%%%%%%%%%%%%%%%%%%%%%%%
\begin{figure}[tbh]
	\subfigure{
		\hspace{-0mm}\includegraphics[width=8cm]{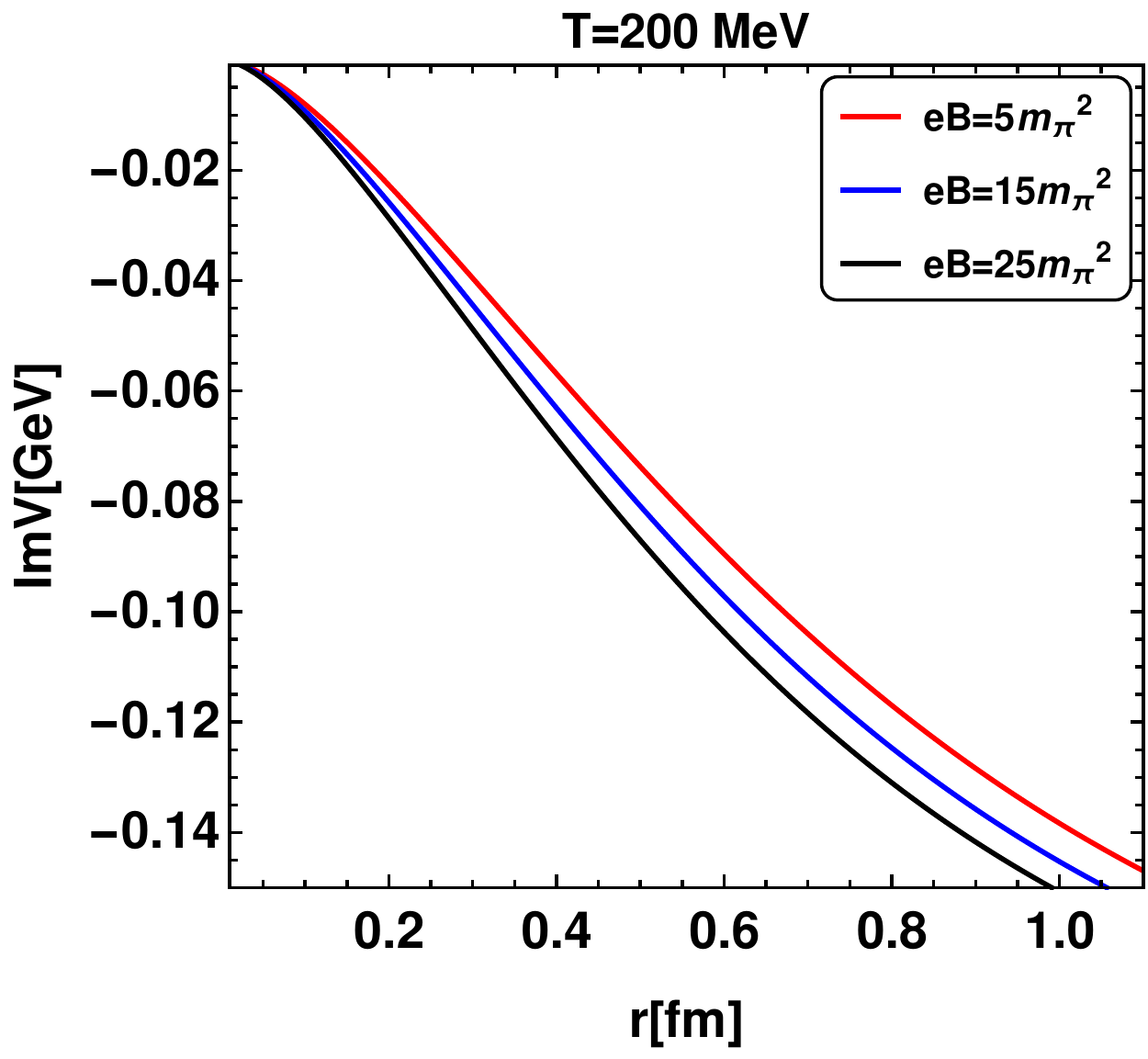}} 
	\subfigure{
		\includegraphics[width=8cm]{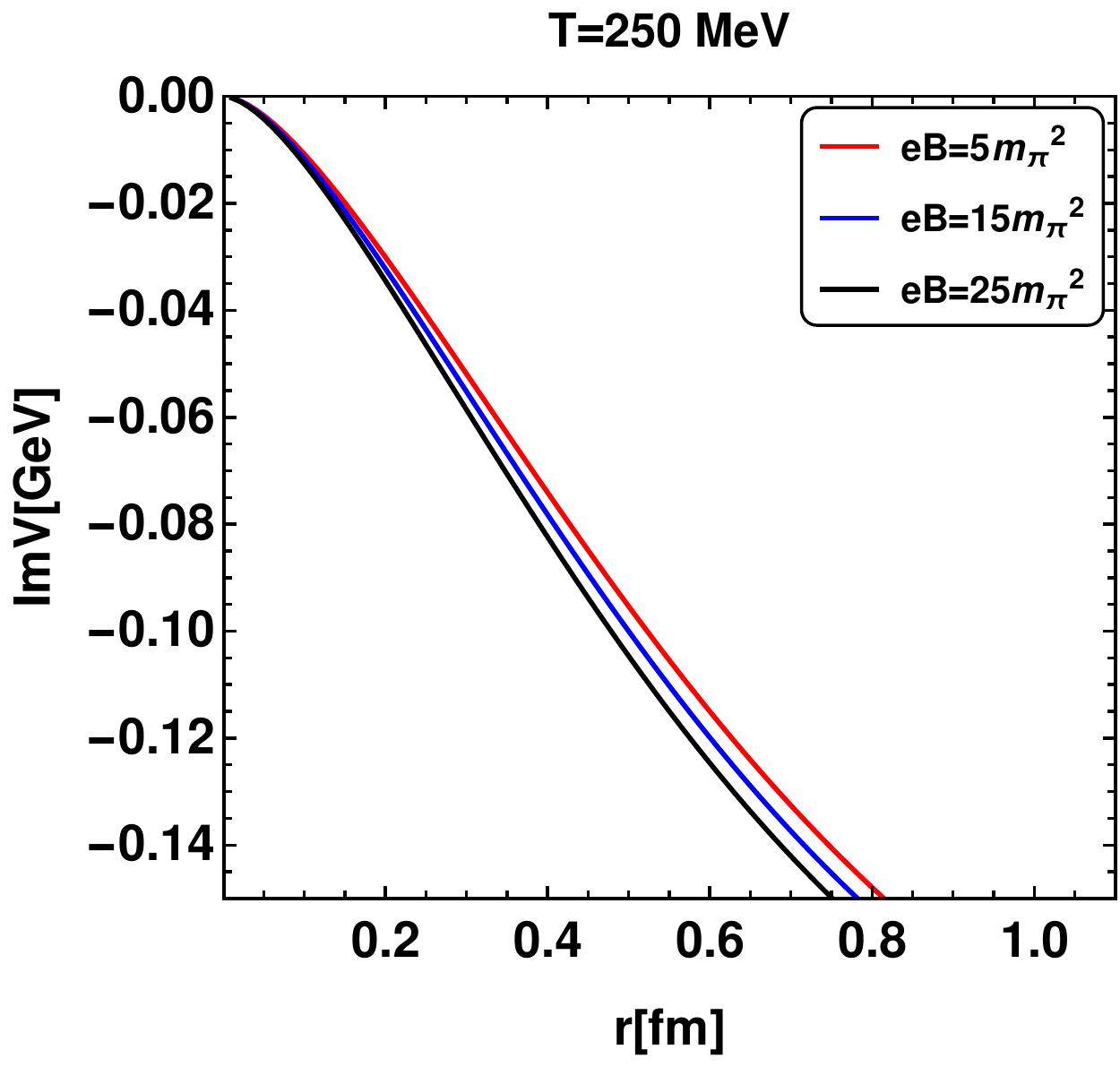}}
	\caption{
	Variation of the imaginary part of the potential with the separation distance, $ r $ for various values of magnetic field 
	at $T=200$ MeV (left) and $T=250$ MeV (right).}
	\label{fig3}
\end{figure}
Figure~\ref{fig3} shows the variation of the imaginary part of the potential with the separation distance ($ r $) for various values of magnetic field ($ ~eB= 5 m_{\pi}^2,~ 15 m_{\pi}^2, ~25 m_{\pi}^2~$) at temperatures $T=200$ MeV and $T=250$ MeV. We find that the imaginary part of the potential increases in magnitude with the increase in magnetic field and hence it
contributes more to the Landau damping induced thermal width obtained from the imaginary part of the potential. The increase in the magnitude of the imaginary part of the potential is more at a higher temperature ($T=250$ MeV) as compared to a lower temperature ($T=200$ MeV) for a given $ r $.

%%%%%%%%%%%%%%%%%%%%%%%%%%%
\section{Decay Width}\label{III}
\label{sec:gamma}
The decay width ($\Gamma$) can be calculated from the imaginary part of the potential. The following formula gives a good approximation to the decay width of $ Q\bar{Q} $ states~\cite{Thakur:2013nia,Thakur:2016cki,Patra:2015qoa}
\begin{equation}
 \Gamma =- \int d^3{\bf r} \,  |\psi({\bf r})|^2\, 
\Im \, 
V(\mathbf{r},T, B) 
%\label{gamma_exp0}
\label{gamma_def}
\end{equation}
where $ \psi ({\bf r} $) is the Coulombic wave function for the ground state and is given 
by
\begin{equation}
\psi({\bf r}) = \frac{1}{\sqrt{\pi a_0^{3}}} e^{-r/a_0},
\label{eq:wavefunction}
\end{equation}
where $ a_0 =2/(m_Q\alpha)$ is the Bohr radius of the heavy quarkonium system. We use the Coulomb-like wave functions to
determine the width
%even though the imaginary part of the potential is not purely Coulombic but 
since the leading
contribution to the potential for the deeply bound quarkonium states in a plasma is Coulombic.
%%%%%%%%%%%%%%%%%%%%%%%%%%%%%%%%%%%%%%%%%%%%%%%%%%%%%%%%%%%%%%%%%%%%%%%%%%%%%%%Figure4%%%%%%%%%%%%%%%%%%%%%%% 
\begin{figure}[tbh]
	\subfigure{
		\hspace{-0mm}\includegraphics[width=8cm]{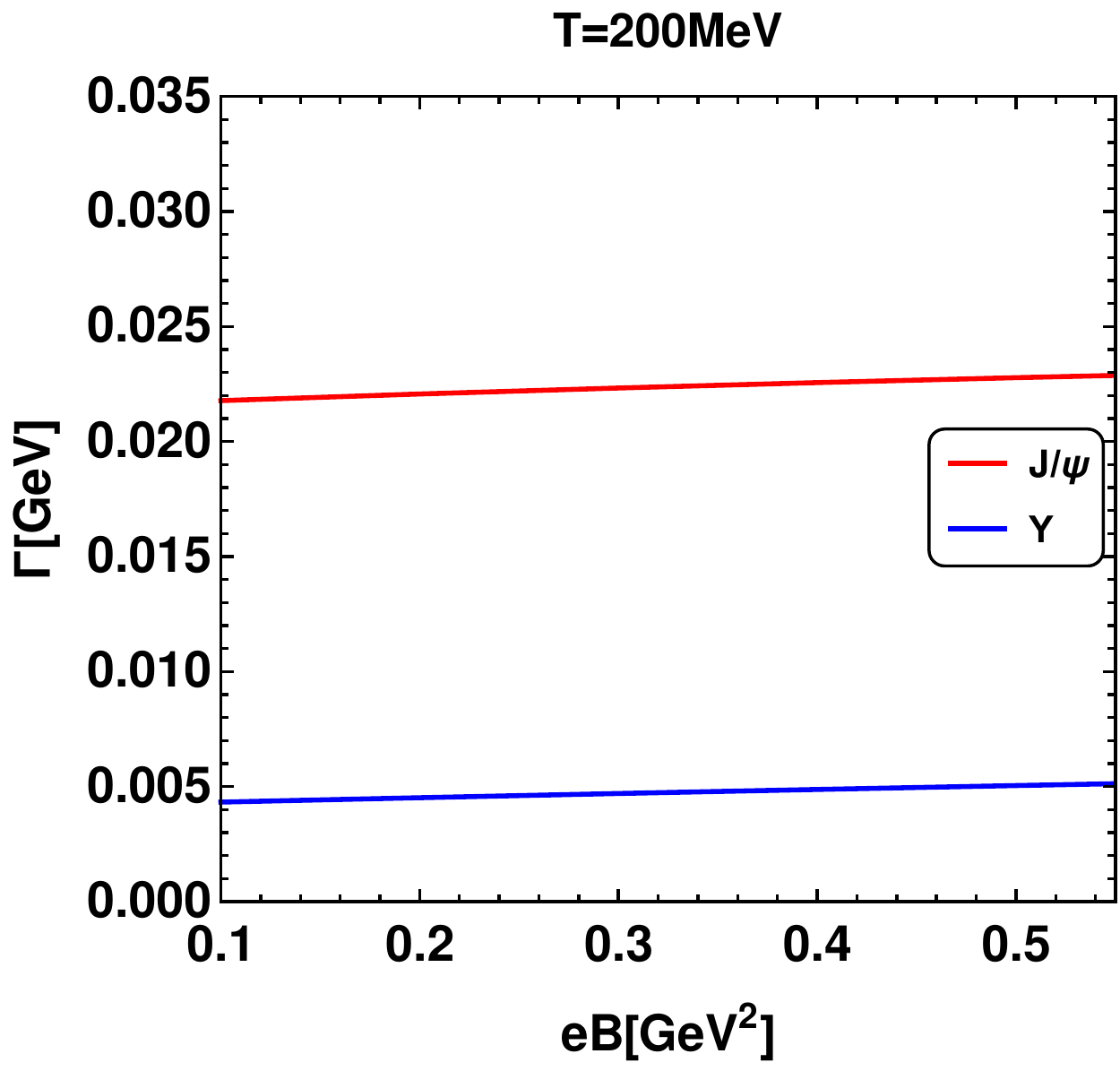}} 
	\subfigure{
		\includegraphics[width=8cm]{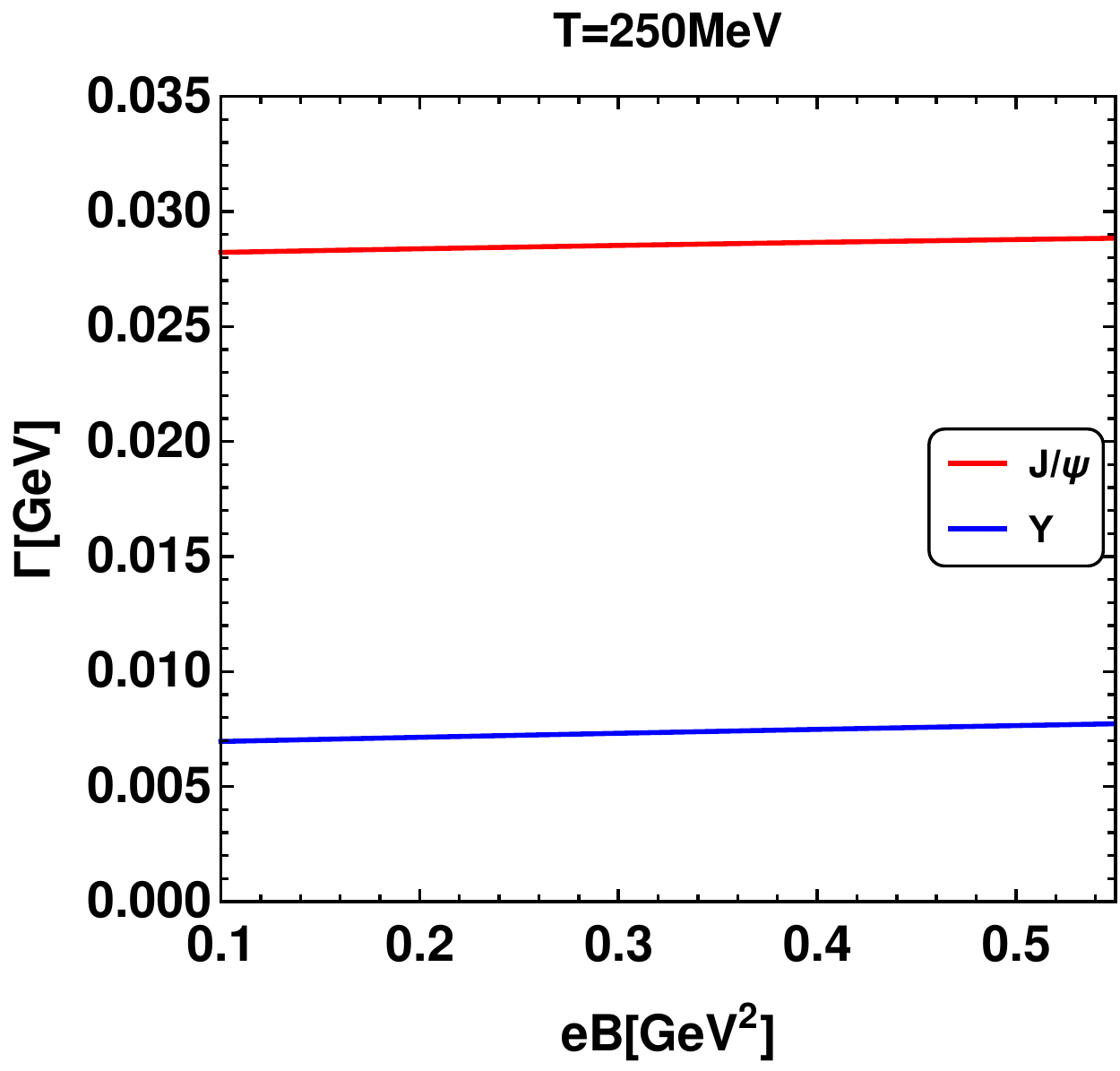}}
	\caption{
		Variation of decay width with the magnetic field for $J/\psi$ and $\Upsilon$ at $ T= 200 $ MeV (left) and $ T=250 $ MeV (right)}.
	\label{fig5}
\end{figure}
%%%%%%%%%%%%%%%%%%%%%%%%%%%%%%%%%%%%%%%%%%%%%%%%%%%%%%%%%%%%%%
\begin{figure}
	\includegraphics[scale=0.65]{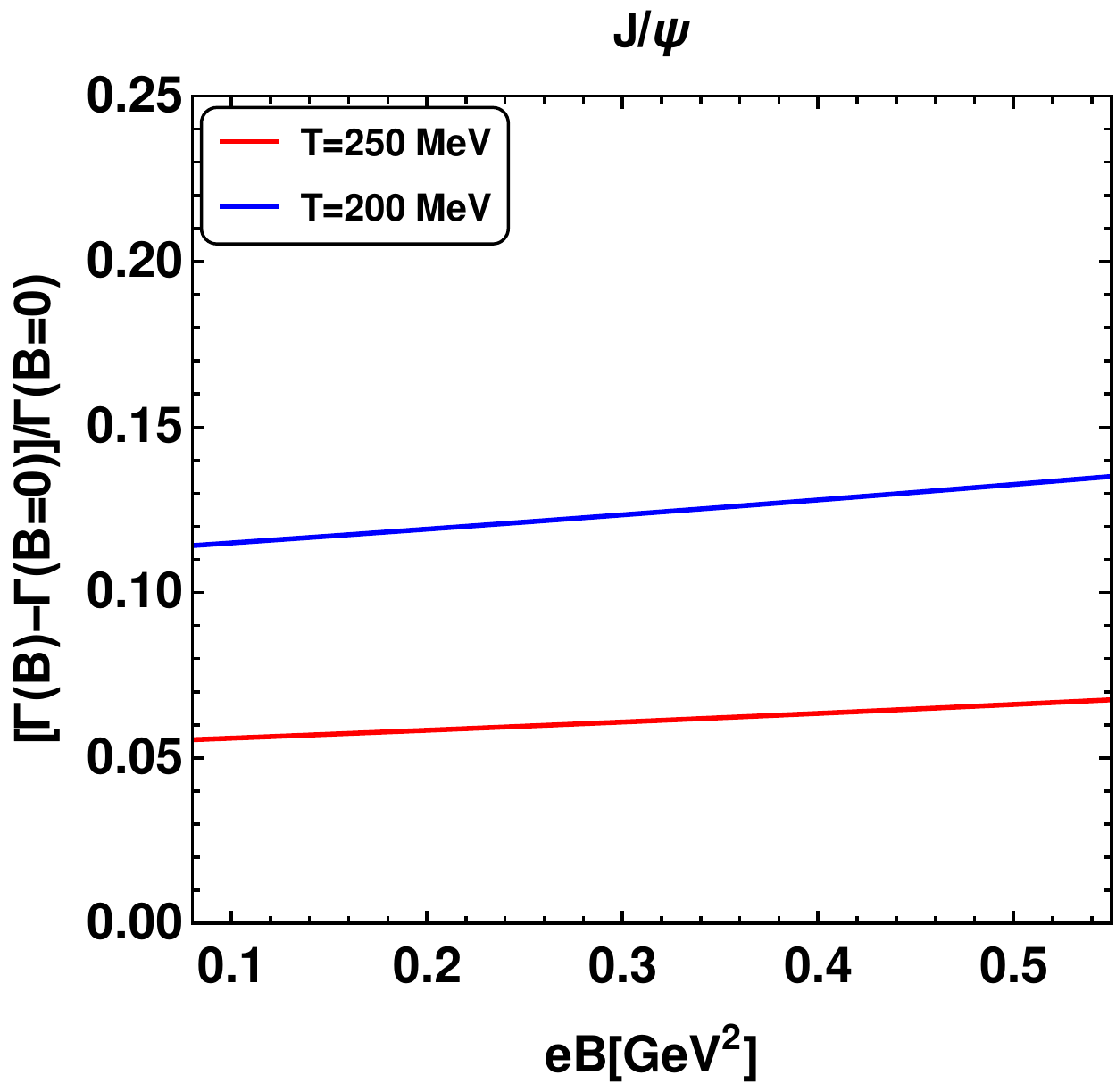}
	\caption{Variation of the ratio of decay width $ [\Gamma(B)-\Gamma(B=0)/\Gamma(B= 0)] $ with magnetic field for $J/\psi$  at $T=200 $ and $ T=250 $}.
		\label{gammavsT}
\end{figure}
 
After substituting the expression for the imaginary part of the potential as given in Eq.~(\ref{imagpt}), in Eq.~(\ref{gamma_def}) we estimate numerically the decay width for given $ T $ and $ B $. In Fig.~\ref{fig5} we show the variation of decay width
for the ground states of charmonium and bottomonium at $ T= 200 $ MeV (left panel) and $ T=250 $ MeV (right panel). 
Here we take 
charmonium and bottomonium masses as $m_c=1.275$ GeV and $m_b=4.66$ GeV respectively 
from Ref.~\cite{Agashe:2014kda}.
We find that the thermal width increases with the increase in magnetic field.
The width for the $\Upsilon$ is
much smaller than the $J/\psi$ 
because the bottomonium states are smaller in size and larger in masses than the charmonium states and hence get dissociated at higher temperatures. 
The width at a higher temperature ($ T=250 $ MeV) is more as compared to a lower temperature ($ T=200 $ MeV) for both $J/\psi$ and $\Upsilon$. 
 Alternatively, we can say that the $ \Gamma $ increases with the increase in magnetic field which results in the early dissociation of quarkonium states.
 
Figure.~\ref{gammavsT} shows the variation of the ratio of the decay width $ [\Gamma(B= 0)-\Gamma(B=0)/\Gamma(B= 0)] $ with magnetic field at  $T=200$ and $T=250$ for $J/\psi$. 
From the figure it is clear that the change in  decay width for $ T=200 $ MeV is about  
$ 11\% $ for $eB \simeq 5 m_\pi^2 $ to as large as $ 14\% $ for $eB \simeq 25 m_\pi^2 $.
For $ T=250 $, the change is from $ 5\% $ to $ 7\% $ in the same range of magnetic field.
The magnetic field effects become weaker at a high temperature as compared to a low temperature. 
Also, we are considering the strong field LLL approximation, i.e., $eB \gg T^2$. This approximation may not hold good at a high temperature.
		
\section{Conclusion}\label{IV}
In this work we have studied the effect of a strong magnetic field on the heavy quark complex potential in a thermal QGP medium. In order to incorporate the
magnetic field effect in the heavy quark potential we first calculated the gluon self-energy in the imaginary time formalism by using the Schwinger propagator in the presence of the magnetic field and then calculated the Debye screening mass ($ m_D $). We have shown the variation of Debye mass with magnetic field and temperature for massless fermions. We found that Debye screening increases in a hot magnetized medium. We have taken only the LLL contribution in our calculation. This is a reasonable approximation as quarkonia are produced during the initial stages of collision when
the magnetic field is very high and the higher Landau levels are at infinity as compared to LLL due to which the LLL dominates and the dimensional
reduction takes place. Further, we study the magnetic field effects on the heavy quark complex potential. This is attempted here through estimating the permittivity of the medium using imaginary time formulation. The presence of magnetic field results is an anisotropic contribution to the gluon propagator from the quark loop. However, such an effect vanishes in the limit of massless quarks and the potential remains isotropic even in the presence of the magnetic field. The effect of the magnetic field in the LLL approximation is manifested through the Debye mass.  
The heavy quark complex potential 
is obtained by bringing together the generalized Gauss law with the characterization
of in-medium effects, i.e., finite temperature and magnetic field,  through the dielectric permittivity. Because of the heavy quark mass ($ m_Q $ ), the requirement
$m_Q \gg T $ and  $m_Q \gg \sqrt{eB} $ is satisfied for the range of magnetic field $ eB=5-25~ m_\pi^2 $.
We have shown the effect of magnetic field on the Debye screening and Landau damping induced thermal
width obtained from the imaginary part of the quark-antiquark potential. In order to see the magnetic field effect on the screening we first show the effect of magnetic field on the real part of the potential.
We found that the real part of the potential
decreases with increase in magnetic field and becomes more screened. The screening also increases with the increase in temperature.
Since the $ Q\bar{Q}$ potential is effectively more
screened in the presence of magnetic field which results in the earlier dissociation of quarkonium states in a strongly magnetized hot QGP medium.

We have also shown the effect of magnetic field on the imaginary part of the potential and hence on the thermal width. The imaginary part of the potential increases in magnitude with the increase in magnetic field and temperature. As a result, the width of the quarkonium states ($ J/\psi $ and $ \Upsilon $ ) get more broadened with the increase in magnetic field and results in the earlier dissociation of quarkonium states in the presence of magnetic field. The width for $ \Upsilon $ is much
smaller than the $ J/\psi $ because bottomonium states are tighter than the charmonium state and hence get dissociated at a higher temperature. 
We found a change in decay width from (11-14)$\%$ at $T=200$ MeV and from  (5- 7)$\% $ at $ T=250 $, for the magnetic field ranging from (5-25)~$m_{\pi}^2$. 
 
The magnetic field effects become very less at a high temperatures; this may be because of the LLL approximation, i.e., $eB \gg T^2$ and this approximation may not be a good approximation at a high temperature.
 Combining both the effects of screening and the
 broadening due to damping, we expect a lesser
binding of a $ Q\bar{Q}$ pair in a strongly magnetized hot QGP medium.

Clearly, the present investigation is limited to very high magnetic field compared to temperature where only lowest Landau level contributes to the dielectric function. For moderate magnetic field one has to include the effects from higher Landau levels. Such an investigation is in progress and will be represented elsewhere.

\section{Acknowledgements}
L.T. would like to thank Najmul Haque for useful comments. We would also like to thank Jitesh Bhatt and Namit Mahajan for useful comments.

%\newpage

\end{document}